\documentclass[prl,twocolumn,superscriptaddress,showpacs]{revtex4}
\usepackage{times}
\usepackage{amsmath}
\usepackage{amsthm}
\usepackage{amssymb}
\usepackage{amsbsy}
\usepackage{graphicx}
\usepackage{pstricks}
\usepackage{color}

\begin{document}

\newcommand{\hexa}{\;
\pspicture(0,0.1)(0.35,0.6)
\psset{unit=0.75cm}
\pspolygon(0,0.15)(0,0.45)(0.2598,0.6)(0.5196,0.45)(0.5196,0.15)(0.2598,0)
\psset{linewidth=0.12,linestyle=solid}
\psline(0,0.15)(0,0.45)
\psline(0.2598,0.6)(0.5196,0.45)
\psline(0.5196,0.15)(0.2598,0)
\endpspicture\;}

\newcommand{\hexb}{\;
\pspicture(0,0.1)(0.35,0.6)
\psset{unit=0.75cm}
\psset {linewidth=0.03,linestyle=solid}
\pspolygon[](0,0.15)(0,0.45)(0.2598,0.6)(0.5196,0.45)(0.5196,0.15)(0.2598,0)
\psset{linewidth=0.12,linestyle=solid}
\psline(0.2598,0.6)(0,0.45)
\psline(0.5196,0.12)(0.5196,0.45)
\psline(0.2598,0)(0,0.15)
\endpspicture\;}

\newcommand{\smallhex}{ \;
\pspicture(0,0.1)(0.2,0.3)
\psset{linewidth=0.03,linestyle=solid}
\pspolygon[](0,0.0775)(0,0.225)(0.124,0.3)(0.255,0.225)(0.255,0.0775)(0.124,0)
\endpspicture
\;}

\begin{abstract}
Excitations which carry ``fractional'' quantum numbers 
are known to exist in one dimension in polyacetylene, and in two dimensions, 
in the fractional quantum Hall effect.   Fractional excitations have also  
been invoked to explain the breakdown of the conventional theory of  
metals in a wide range of three-dimensional materials.   However the  
existence of fractional excitations in three dimensions remains highly controversial.      
In this Letter we report direct numerical evidence for the existence of an extended quantum 
liquid phase supporting fractional excitations in a concrete, three-dimensional  
microscopic model --- the quantum dimer model on a diamond lattice.
We demonstrate explicitly that the energy cost of separating 
fractional monomer excitations
vanishes in this liquid phase, and that its energy spectrum 
matches that of the Coulomb phase in $(3+1)$ dimensional quantum electrodynamics. 
\end{abstract}

\title{A quantum liquid with deconfined fractional excitations in three dimensions}

\author{Olga Sikora}
\address{Max-Planck-Institut f{\"u}r Physik komplexer Systeme, 01187 Dresden, Germany}

\author{Frank Pollmann}
\address{Department of Physics, University of California, Berkeley, CA94720, USA.}

\author{Nic Shannon}
\address{H. H. Wills Physics Laboratory, University of Bristol, Tyndall Avenue, Bristol BS8 1TL, UK.}

\author{Karlo Penc}
\address{Research Institute for Solid State Physics and Optics, H-1525 Budapest, P.O.B. 49, Hungary.}

\author{Peter Fulde}
\address{Max-Planck-Institut f{\"u}r Physik komplexer Systeme, 01187 Dresden, Germany}
\address{Asia Pacific Center for Theoretical Physics, Pohang, Korea}

\maketitle

One of the great triumphs of twentieth-century physics was to show how the different physical
properties of metals, magnets, semiconductors and superconductors could be understood in terms of the 
collective properties of a single elementary particle --- the electron.  
For over fifty years, Landau's concept of the Fermi liquid, a three-dimensional quantum liquid whose quasi-particle excitations 
carry the same spin and charge quantum numbers as an electron, has served as the ``standard model'' for metals.   
Recently, however, this theory has been challenged by experiments on a wide range of strongly-correlated materials, 
including quasi-one dimensional conductors, cuprate high-temperature superconductors
and heavy Fermion systems near a quantum critical point~\cite{Schofield99,varma02}.

These experiments prompt us to question whether new types of quantum liquid, capable of supporting new types of excitation 
might exist in nature?   Indeed, excitations with fractional charge are known to exist in highly-doped, one-dimensional, trans-polyacetylene~\cite{Su81}, 
and in the two-dimensional fractional quantum Hall effect~\cite{Laughlin83}.   
However the existence of fractional excitations in three dimensions remains highly controversial. Indeed, since the understanding the quantum Hall 
effect is bound to the two-dimensional concept of ``anyons'' with fractional statistics~\cite{Wilczek82}, it has often been argued to be impossible.

An unambiguous way of resolving this question would be to find unbiased evidence of the existence a quantum liquid 
supporting deconfined fractional excitations in a concrete, three-dimensional microscopic model.   This is the goal of this Letter.   
The model we consider is the quantum dimer model 
 \begin{eqnarray}
H=&-&\sum_{\{\smallhex\}}\left(\Big|\hexa\Big\rangle\Big\langle\hexb\Big|+\mbox{H.c.}\right)\nonumber\\
&+&\mu\sum_{\{\smallhex\}}\left(\Big|\hexa\Big\rangle\Big\langle\hexa\Big|+\Big|\hexb\Big\rangle\Big\langle\hexb\Big|\right).
\label{eq:QDM}
\end{eqnarray}
on a three-dimensional diamond lattice, where the first term describes the kinetic energy of hard-core dimers 
tunneling between different degenerate configurations on a hexagonal ``flippable'' plaquette, 
and $\mu$ sets the ratio of potential to kinetic energy.    This model was recently derived as an effective description 
of half-magnetization plateaux in Cr spinels~\cite{Bergman2006-PRL,Bergman2006-PRB}, 
and also describes spin-less fermions or hard-core bosons on a 
pyrochlore lattice at quarter filling, in the limit of strong nearest-neighbour interactions~\footnote{In this special case, the Fermi sign  can be ``gauged'' out of the problem by a suitable choice of convention for labeling lattice sites.}. 
\begin{figure}[tbp]
\begin{center}
\includegraphics[width=7cm]{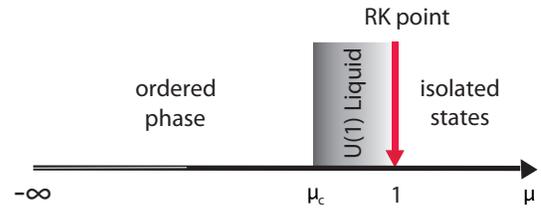}
\end{center}
\caption{(Color online) Conjectured form for the ground state phase diagram of the quantum dimer model 
on a bipartite lattice in 3D, as a function of the ratio $\mu$ of potential to kinetic energy, 
following~\protect\cite{Moessner03b,Bergman2006-PRL,Bergman2006-PRB}.} 
\label{fig:phase-diagram}
\end{figure}
 
In fact quantum dimer models arise naturally as effective models of many different condensed matter systems, 
and provide a concrete realizations of several classes of lattice gauge 
theory~\cite{Rokhsar88,FradkinBook,Moessner01a,Diep_book2004,Trousselet2008}.   
As such, they have become central to the theoretical search for new quantum phases and excitations.   
A key feature of these models is the existence of a gapless ``Rokhsar-Kivelson'' (RK) point
for $\mu=1$, at which all correlation functions exhibit algebraic decay~\cite{Rokhsar88}.  
Doping the model by removing a dimer introduces two monomers.   
Precisely at the RK point these monomer excitations are independent, deconfined excitations, each carrying 
half of the spin/mass/charge associated with a single dimer.    

Field theory arguments suggest that, on bipartite lattices in three dimensions, 
 the RK point ``grows'' into an extended quantum liquid phase
 --- see Fig.~\ref{fig:phase-diagram}.   
Similar behaviour is expected in a class of closely related quantum loop models \cite{Fulde02,Hermele04a, Banerjee08}.   
In both cases the liquid phase is described by the $U(1)$ gauge theory corresponding to the Coulomb phase in 
\mbox{$(3 + 1)$}-dimensional quantum electrodynamics~\cite{Huse03, Moessner03b, Bergman2006-PRB,Hermele04a}.   
Within this effective field theory, monomers are {\it deconfined} magnetic monopoles, sometimes referred to as spinons, 
whose mutual interactions fall off as $~1/r^2$~\footnote{For classical analogues to this problem, 
see e.g. C. Castelnovo {\it et al.}, Nature {\bf 451} 42, (2008); G. Misguich {\it et al.} Phys. Rev. B {\bf 78}, 100402(R) (2008)}.   
Thus, if we can establish the existence of a $U(1)$ liquid phase in the 
quantum dimer model 
defined by Eq.~(\ref{eq:QDM}), we will have found a concrete route to stabilizing fractional excitations in three dimensions.  

In this Letter we use a range of numerical techniques to determine the phase diagram of the quantum dimer model 
on a diamond lattice {\it directly} from its microscopic Hamiltonian Eq.~(\ref{eq:QDM}).   
Our results explicitly confirm the conjectured form of the phase diagram shown in Fig.~\ref{fig:phase-diagram},  
affirming the existence a $U(1)$ liquid phase and the absence of a confining string potential for 
fractional monomer (monopole) excitations.

\begin{figure}[tbp]
\begin{center}
\includegraphics[width=7.0cm]{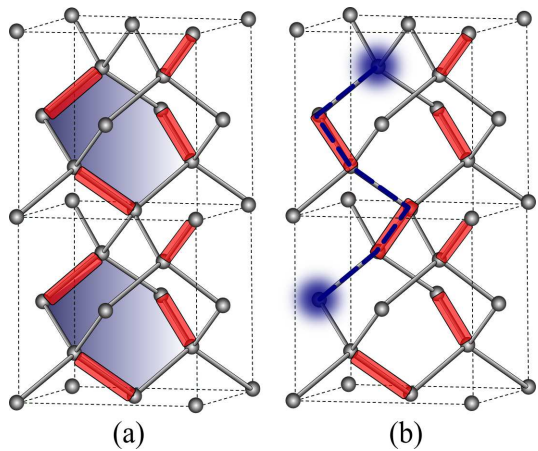}
\end{center}
\caption{(Color online) (a) Two adjacent 16-bond cubic unit cells of the diamond lattice, showing a dimer
configuration in the maximally-flippable {\sf R}-state.  The two flippable hexagons contained within this picture are 
shaded blue.  (b) Non-local ``string'' defect (dashed line) created by separating two monomer excitations.}
\label{fig:diamond}
\end{figure}

We begin by considering the ordered phase for \mbox{$\mu\rightarrow-\infty$}.    Here the potential energy dominates, 
and the ground state of Eq.~(\ref{eq:QDM}) is the set of dimer configurations which maximize the number of 
``flippable'' hexagons, with one out of four hexagons being flippable.   This is the so-called {\sf R}-state \cite{Bergman2006-PRB}, illustrated in Fig.~\ref{fig:diamond}(a), which has cubic symmetry and is eight-fold degenerate.     To establish 
the validity of the proposed phase diagram Fig.~\ref{fig:phase-diagram}, we need to connect this with the 
RK point $\mu=1$, for which the 
ground state the equally weighted sum of all possible dimer configurations~\cite{Rokhsar88}.   
We accomplish this using a mixture of exact diagonalization, 
variational Monte Carlo (VMC) and Green's function Monte Carlo (GFMC) techniques.  

GFMC is a zero temperature quantum Monte Carlo technique which offers 
a systematic way of improving upon the variational wave function output by a variational Monte Carlo calculation~\cite{Trivedi1990, Calandra1998}.  Where it converges, GFMC offers accuracy comparable with exact diagonalization~\footnote{We have checked our 
GFMC results explicitly against exact diagonalization for small system sizes, and for large systems sizes against an expansion in $1/|\mu|$ for $\mu \to - \infty$, and perturbation theory in $1-\mu$  for $\mu \to 1$.   
We find complete agreement in all cases.   This (lengthly) analysis will be presented elsewhere.}.  
As an input for VMC, we use a trial wave function based on hexagon-hexagon correlations, with of order $40$ variational parameters.
Cluster sizes are limited by the rapid growth of the Hilbert space with system size --- there are $~1.3^{N}$
dimer coverings of an $N$-site diamond lattice cluster, and {\it all} of these 
contribute to the ground state wave function approaching the RK point.   
However we are able to access families of clusters with edges parallel to the $[100]$, $[110]$ and $[111]$ 
directions which have the full (cubic) symmetry of the diamond lattice.    Below we focus mainly on $[100]$ 
clusters with $2 L^3$ diamond lattice bonds, where $L=\{4, 6, 8, 10\}$.

\begin{figure}[tbp]
\begin{center}
\includegraphics[width=8.0cm]{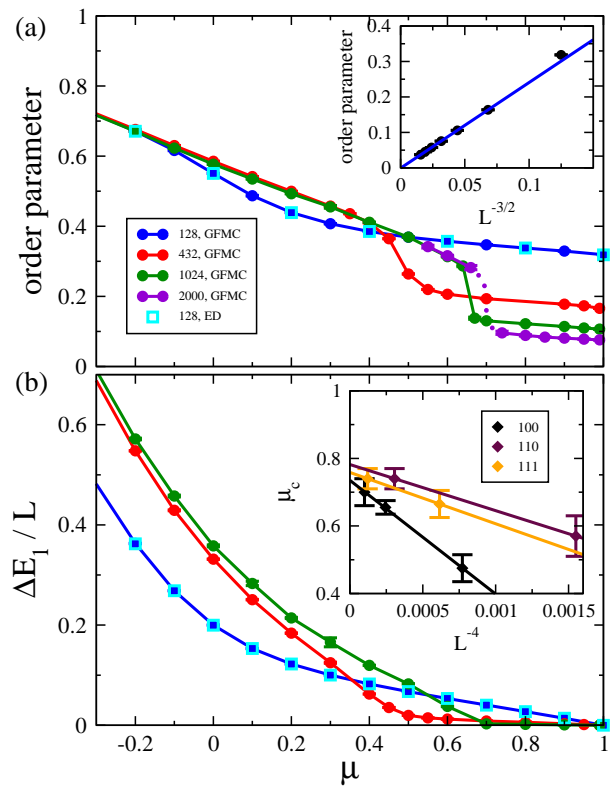}
\end{center}
\caption{(Color online) (a) Order parameter $ m_{\sf R}$ of the {\sf R}-state 
as a function of $\mu$, calculated 
using Green's function Monte Carlo (GFMC) for $[100]$ clusters with 128, 432, 1024 and 2000 bonds
(lines and dots serve as a guide for the eye).   
Results are normalized such that $ m_{\sf R} =1$ for $\mu \rightarrow -\infty$.  
Exact diagonalization (ED) results for 128 bonds are also shown.  
A sharp jump in $ m_{\sf R} $ can be seen for 
$\mu_c \approx 0.5$--$0.7$, suggesting a first order transition out of the {\sf R}-state.   
(b) This coincides with a collapse in the string tension $\Delta_1/L$.   
The inset in panel (a) shows the finite-size scaling of $ m_{\sf R}$ at the RK point ($\mu=1$).
The inset in panel (b) shows the finite-size scaling of $\mu_c$.  We obtain a value $\mu_c = 0.77\pm0.02$ }
\label{fig:order_and_string}
\end{figure}

In Fig.~\ref{fig:order_and_string}(a) we present numerical results for the order parameter $m_{\sf R}$.   
This is defined as a sum of projections onto the six linearly-independent combinations of the eight degenerate ${\sf R}$
states.  For the smallest 128-bond cluster, $m_{\sf R}$ evolves smoothly with $\mu$.  
There is a perfect numerical agreement between GFMC and exact diagonalization results, and very close agreement between these and VMC.   
For larger systems there is a strong suggestion of a {\it first order} transition out of the {\sf R}-state --- a jump in $m_{\sf R}$ is observed for 
$\mu_c \approx 0.7$ in GFMC simulations (2000 bonds).   We have studied the finite size scaling of $\mu_c$ for series of
$[100]$, $[110]$ and $[111]$ clusters, and find $\mu_c=0.77\pm0.02$ in the thermodynamic limit --- see inset to Fig.~\ref{fig:order_and_string}(b)~\footnote{The $L^{-4}$-scaling follows from the fact that the {\sf R}-state is gapped, while the competing liquid phase has linearly dispersing photon excitations.}.  
For the clusters considered, $m_{\sf R}$ still takes on a finite value for $\mu > \mu_c$, but this is a finite-size effect.   
Exactly at the RK point $m_{\sf R}$ {\it must} vanish, and we can use our knowledge of the exact ground state to simulate 
much larger clusters.   We find that that $m_{\sf R}$ scales to zero as $L^{-3/2}$ (see the inset to Fig.~\ref{fig:order_and_string}(a)). 

 
Another indicator of a crystalline, ordered phase is the linear, confining ``string potential'' associated 
with separating two fractional monomer excitations --- illustrated in Fig.~\ref{fig:diamond}(b).   
Conversely, the energy cost of separating two fractional excitations by a distance $L$ must {\it vanish} in the quantum liquid 
we are seeking.     A suitable ``string'' configuration can be prepared by joining the ends of the string defect 
shown in Fig.~\ref{fig:diamond}b) such that it connects opposite faces of a (periodic) cluster of linear dimension $L$.   
Since GFMC preserves quantum numbers, it can be used 
to calculate the energy of the quantum eigenstate corresponding to such a classical string configuration, 
and the ``string tension'' calculated as $\Delta_1/L$, where $\Delta_1$ is the energy of the string 
excitation relative to the ground state.

Deep inside the ordered state,  $\Delta_1$ scales as $\propto \mu L$, 
however the string tension $\Delta_1/L$ collapses abruptly at the value of $\mu$ for which
the order parameter jumps --- see main panel, Fig.~\ref{fig:order_and_string}(b). 
Taken together, these facts are strongly suggestive of a first order transition from
a crystaline to a liquid phase with deconfined monomer excitations. 
But they do not yet prove the existence of the $U(1)$ liquid we are looking for.
In order to accomplish this, we must test explicitly the predictions of the relevant 
effective field theory.   

The defining property of a dimer model is that every lattice site is touched by {\it exactly} one dimer.   
On a bipartite lattice, we can implement the this constraint by associating a magnetic field
$\mathbf B=\mathbf{\nabla} \times \mathbf A$ 
with each dimer and each empty bond.
This field points from site to 
site along each bond 
such that $\mathbf\mathbf{\nabla}\cdot \mathbf B=0$ at {\it every} diamond lattice site~\cite{Huse03,Moessner03b,Hermele04a}.   
The total flux $\phi = \int d {\mathbf S}\cdot{\mathbf B} $ through any plane (cutting bonds)
in the lattice is a conserved quantity.   
Therefore, for periodic boundary conditions, the flux through a set of orthogonal planes defines a set of (integer) topological 
quantum numbers $(\phi_1,\phi_2,\phi_3)$.    We choose to work in a gauge where $\mathbf{\nabla}\cdot\mathbf{A}=0$   
and in this Letter consider only flux sectors of the type $(\phi,0,0)$, where $\phi=1$ denotes the smallest finite
flux and corresponds to the state with a single string defect described above.

This representation clearly has a lot in common with conventional electromagnetism  
and, following~\cite{Moessner03b},  we can use this analogy to write down a plausible long-wavelength 
action for the QDM on a diamond lattice
\begin{equation}
\mathcal{S}=\int d^3xdt\left[ \mathbf E^2 -\rho_2 \mathbf B^2 -\rho_4(\mathbf{\nabla \times \mathbf B})^2 
\right],
\label{eq:QDMAC}
\end{equation}
where $\mathbf E=\partial_t\mathbf A -\mathbf{\nabla}A_0$ and $\rho_4 >0$.  
We have studied $\rho_2$ in a perturbation theory about the RK point and find that it varies as
$\rho_2 \sim 0.6(0) \times (1-\mu) $.    
For  $\mu > 1$, the coefficient $\rho_2  < 0$, and the system chooses 
those dimer configurations with the maximum possible flux $\phi$  --- a set of isolated states which are 
not connected by {\it any} matrix element of the Hamiltonian Eq.~(\ref{eq:QDM}).   All excitations are then gapped.
At the RK point $\rho_2$ vanishes, and all flux sectors contribute equally to the ground state. 
The system possesses gapless, transverse excitations with dispersion $\omega = \rho_4 k^2$.  
However for $\mu \lesssim 1$, $\rho_2  > 0$ and the analogy with electromagnetism is complete 
--- transverse excitations are now``photons'' with dispersion $\omega = \sqrt\rho_2|k|$~\footnote{We work in units such that the volume of the
16-bond cubic unit cell is equal to 8.   This, together with the normalization chosen for the the flux $\phi$, defines
the units of $\rho_2$ and therefore the speed of light for photon excitations.}.  

These photons are the signature feature of the proposed $U(1)$ liquid state, and offer a beautiful 
realization of Maxwell's laws in a condensed matter system.   However, 
Eq.~(\ref{eq:QDMAC}) also contains information about the {\it finite size}
scaling of the ground state energy, as a function of flux $\phi$.   
A flux $\phi$ through a cluster of volume $L^3$ corresponds to an average magnetic 
field $B = \phi/L^2$.  In the $U(1)$ liquid state, this magnetic field is uniformly distributed on the ``coarse-grained'' 
scale of the effective action Eq.~(\ref{eq:QDMAC}).  It then follows from Eq.~(\ref{eq:QDMAC}) that the energy difference 
$\Delta_\phi = E_\phi - E_0$ between the ground state of the zero flux sector, and 
the lowest energy state of the sector with flux $\phi$ scales as
\begin{equation}
\label{EQN:Delta}
\Delta_ \phi = E_\phi - E_0 = \rho_2 \frac{\phi^2}{L}\ .
\end{equation}
Furthermore, by cycling dimers across the boundaries of the cluster, we can systematically construct 
the different flux sectors $\phi$.  This is the natural generalization of the ``thought experiment'' measuring
the string tension $\Delta_1/L$, above.    Combining these results, we have a systematic means 
of studying the spectroscopic signatures of the proposed $U(1)$ liquid state in simulations of a finite size systems.

\begin{figure}[tbp]
\begin{center}
\includegraphics[width=8.0cm]{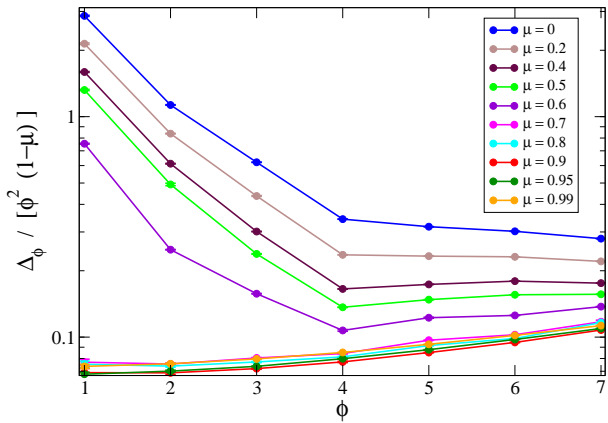}
\end{center}
\caption{(Color online) Energy gap $\Delta_\phi$ normalized to the square of the ``magnetic'' flux $\phi$ and distance from RK point, 
$1 - \mu$, plotted on a log-linear scale for values of $\mu$ spanning the ordered {\sf R}-state and the proposed $U(1)$ liquid phase.      
In a $U(1)$ liquid, $\Delta_\phi/\phi^2 \sim {\text const.}$, while deep within in the ordered {\sf R}-state, $\Delta_\phi/\phi^2 \sim 1/\phi$.   
Here, a clear division is observed between $\mu \ge 0.7$ (liquid) and $\mu \le 0.6$ (ordered).  
Results are from Green's function Monte Carlo (GFMC) calculations for a cluster of 1024 bonds.}
\label{fig:excitations}
\end{figure}

In Fig.~\ref{fig:excitations} we present this analysis for GFMC simulations of an $L=8$ cluster of 1024 bonds.   
Deep within the ordered phase 
$\Delta_ \phi^{\sf R} \propto L\times\phi$, 
and a plot of $\Delta_ \phi/\phi^2$ at fixed $L$ should therefore show a clear distinction between
a $U(1)$ liquid ($\Delta_ \phi/\phi^2  \propto \text{const.}$), and an ordered phase ($\Delta_ \phi/\phi^2  \propto 1/\phi$).   
For this cluster size a clear division is observed between $\mu \ge 0.7$ (liquid) and $\mu \le 0.6$ (ordered).
This is unambiguous evidence of a phase transition from a linearly confining phase into a $U(1)$ liquid.   
The abrupt, qualitative change in the spectra indicates that this phase transition is first-order in character.  
Since we already know the nature of the confining phase --- the {\sf R}-state --- 
this completes the phase diagram.

It is interesting to compare these results with recent work on the problem of strongly interacting, hard core bosons at half-filling on a pyrochlore 
lattice~\cite{Banerjee08}.   For large values of nearest-neighbour interaction, this hard-core bosonic model 
can be approximated by an effective quantum-loop model on the diamond lattice, closely related to the quantum dimer model studied here. 
Equal-time and static density correlation functions for the Bosonic model were calculated using a path integral quantum Monte Carlo approach,  
and an insulating  ``liquid''  state was found, whose finite-temperature correlations were well described by the predictions of a 
$U(1)$ gauge theory.  However the excitations of this model have yet to be studied, and it remains an open question whether this liquid 
persists down to zero temperature.

In conclusion, the numerical results presented in this Letter establish the existence of a $U(1)$ liquid ground state
in the quantum dimer model on a diamond lattice for a finite range of parameters $0.7(7) < \mu < 1$, confirming the 
field theoretical scenario proposed  by~\cite{Huse03, Moessner03b, Bergman2006-PRB}.    This liquid state 
has been demonstrated explicitly to support deconfined monomer excitations.   These results therefore confirm 
the existence of fractional excitations in three dimensions.   
 
It is our pleasure to acknowledge helpful conversations with Kedar Damle, Yong-Baek Kim, Gregoire Misguich, Roderich Moessner, and Arno Ralko.  
We are particularly indebted to Federico Becca for advice and encouragement with simulation techniques.   This work was 
supported under EPSRC grants EP/C539974/1 and EP/G031460/1, Hungarian OTKA grants K73455 and K62280, 
U.S. National Science Foundation I2CAM International Materials Institute Award, Grant DMR-0645461, and the guest 
programs of MPI-PKS Dresden and YITP, Kyoto.

\bibliographystyle{apsrev}

\end{document}